\documentclass[showpacs,preprintnumbers,nofootinbib, amsmath,amssymb,twocolumn]{revtex4}
\usepackage{graphicx}
\usepackage{amsmath}
\usepackage{natbib}
\usepackage{aas_macros}

\begin{document}

\title{Nuclear Statistical Equilibrium neutrino spectrum}

\author{Andrzej Odrzywolek}
 \homepage{http://www.ribes.if.uj.edu.pl/}
 \email{odrzywolek@th.if.uj.edu.pl}
\affiliation{
M. Smoluchowski Institute of Physics\\ 
Jagiellonian University\\  
Reymonta 4\\ 
30-059 Krakow\\ 
Poland
}%

\date{\today}

\begin{abstract}

The spectral emission of neutrinos from a plasma in nuclear statistical equilibrium (NSE) is investigated. Particular attention is paid to the possible emission of high energy ($>$10 MeV) neutrinos or antineutrinos. A newly developed numerical approach for describing the abundances of nuclei in NSE is presented. Neutrino emission spectra, resulting from general Fuller, Fowler, Newman (FFN) conditions, are analyzed. Regions of T-$\rho$-$Y_e$ space favoring detectability are selected. The importance of critical $Y_e$ values with zero net rate of neutronization ($\dot{Y_e}$) is discussed. Results are provided for the processing of matter under conditions typical for thermonuclear and core-collapse supernovae, pre-supernova stars, and neutron star mergers.

\end{abstract}

\pacs{97.60.Jd, 26.60.+c, 97.60.-s, 26.30.-k, 26.50.+x}

\maketitle

\section{Introduction}

Neutrino cooling is of paramount importance in the modern astrophysics \cite{Arnett, Bisnovatyi-Kogan, Kippenhahn}.
It governs late stages of stellar evolution, especially massive
stars ($>$8-10 M$_\odot$) \cite{MassiveStars_rev1,Young1}, red giant cores \cite{Raffelt}, white dwarfs \cite{KantorGusakov}, 
core-collapse supernovae 
\cite{2008ApJ...688.1159M, 2008PhRvL.100b1101D, 2006JPhCS..46..393B, 2007ApJ...666.1140N, 2006A&A...447.1049B, 2006A&A...457..281B} 
and (proto)neutron stars \cite{Hansel}. Neutrino emission is important 
in mergers involving neutron star  \cite{RosswogWynn, DessartOtt, 1999ApJ...527L..39J, 2005A&A...436..273A, 2001A&A...380..544R}, 
the dense accretion disks of Gamma Ray Bursts (GRB) models 
\cite{NuCooledDisks1, NuCooledDisks2, 2007A&A...463...51B}, 
type Ia supernovae \cite{KunugiseIwamoto} and X-ray flashes \cite{X-ray}.

Usually, neutrinos carry away energy, and only the total neutrino emissivity, i.e. amount
of energy carried out by neutrinos is of interest. The neutronization induced
by the net $\nu_e-\bar{\nu}_e$ flux is 
crucial for understanding of the nucleosynthesis. Therefore previous research on NSE neutrino emission \cite{Iwo}
focused on: (1) $\nu_e-\bar{\nu}_e$ particle emission rates and
(2) total $\nu_e+\bar{\nu}_e$ energy carried out by the neutrinos.
We would like to extend this analysis to cover spectral/flavor properties of the NSE neutrino
flux. 

In known research a detailed treatment of the neutrino emission is done for core-collapse simulations 
\cite{BurrowsThompsonPairs, Bruenn1985}. On the other hand, it is frequently neglected for other 
astrophysical objects (e.g. Ia supernovae).
Nowadays more interest is dedicated towards spectral properties
of the neutrino flux. The neutrino energy is important for core-collapse supernovae,
for the neutrino-induced nucleosynthesis ($\nu$--process, \cite{NeutrinoNucleosynthesis, nu-p, SN-winds}), 
neutrino oscillations, and for the detection of neutrinos in terrestrial experiments.
The last area is poorly explored.
The neutrino spectrum for neutrino cooling processes
rarely is treated in rigorous way. Typical procedure is to use some
more or less justified analytic forms for the neutrino
energy spectrum. There are parameters that are found from known neutrino
emissivity  and the average neutrino energy. In this paper we continue our former investigation \cite{OMK3, Odrzywolek_plasmaneutrino}
to find spectral properties for important neutrino emission processes, We proceed now to
processes involving weak nuclear $\beta$
transitions.

Neutrino cooling processes can be separated into two classes. There are (1) thermal
processes including $e^- e^+$ pair annihilation, massive in-medium photon
\& plasmon decay and neutrino photoproduction, and (2) weak nuclear processes (i.e.
$\beta^\pm$ decays and $\epsilon^\pm$ captures). 
We would like to point out that for all thermal  processes (pair, plasma, photoproduction, bremsstrahlung,
neutrino de-excitation of the nuclei) the neutronization rate vanishes, i.e.
the change of the proton/neutron ratio is  due exclusively to  weak nuclear processes.
Class (1) produces all flavors
while (2) only $\nu_e$ and $\bar{\nu}_e$. However, neutrino oscillations
can mix flavors. Information on thermal and weak components might be destroyed. 
It happens somewhere between emission and interaction/detection.

We assume that matter is transparent to neutrinos.
Therefore, weak nuclear processes often tend to dominate neutrino emission of hot and very dense
plasma. In particular,
electron captures by both protons and heavy nuclei are progressively more intense.
With growing density, the Fermi energy $E_F \simeq \mu_e$ can become larger than the capture 
threshold (Q-value), 
for increasing number of nuclear species. High temperature additionally
enhances emission. Many of the nuclei remain in the thermally 
excited states. Matrix elements for  these  weak transitions are often large. For temperatures
above $\sim$0.5~MeV, a significant fraction of equilibrium positrons builds up. This causes a strong
$\bar{\nu}_e$ flux due to $e^+$ captures, particularly on free neutrons.

In contrast to thermal processes, determined entirely (including energy spectrum) 
by the local thermodynamic properties of matter (e.g. temperature $kT$
and electron chemical potential $\mu_e$), weak nuclear processes
depend also on abundances of nuclei. This renders the task of calculating
neutrino spectrum difficult to achieve. This is especially true for
evolutionary advanced 
objects\footnote{This situation is however very difficult to describe using statistical 
methods. Variety of astrophysical objects and processes make it closer to complex systems
rather than gases.
}. 
All that we can say for rapidly evolving object, is that the neutrino spectrum
emitted from plasma is of the form:
\begin{equation}
\phi(\mathcal{E}_\nu)  = \sum_k X_k(t) \psi_k(\mathcal{E}_\nu, kT, \mu_e) \;  \frac{\rho}{m_p A_k}.
\label{Xk}
\end{equation}
Here $\psi_k$  represent (assumed known, from theory or experiment) spectral shape of single nuclei
neutrino emission, and $X_k(t)$ set of usually unknown and rapidly varying abundances. 
Tracking of the required
required number of a few hundred abundances  is possible
at most in simplest one-dimensional models, to our knowledge.

Fortunately, if the temperature becomes high enough, nuclei begin
to ,,melt'' due to photo-disintegrations. Nuclei re-arrange
due to strong interactions into the most probable state 
favored by the thermodynamics\cite{Iwo2}.
This is the Nuclear Statistical Equilibrium (thereafter NSE) approximation \cite{CliffordTayler}.
The timescale required to achieve NSE is temperature-dependent \cite{NSE-conditions-rus, NSE-conditions-eng}.
It can be approximated as \cite{NSE-timescale}: 
      \begin{equation}
      \label{time-to-nse}
      \tau_{\mathrm{NSE}} \sim \rho^{0.2} e^{179.7/T_9 - 40.5}
      \end{equation}
where $\rho$ is the density in $g/cm^3$, $T_9=T/10^9$K and $T$ is the temperature in K.
Eq.~\eqref{time-to-nse} provides one of the most important constraints limiting
    the use of the NSE approach. We assume implicitly in \eqref{time-to-nse}
that $Y_e=0.5$ \cite{NSE-conditions-rus, NSE-conditions-eng}. Therefore in a plasma
with the value of $Y_e$ which is far from 0.5 caution is required.
Both under- an over- estimate is possible. The timescale is of the order of
the age of the universe, $\tau_{\mathrm{NSE}}\sim10^9$ years, for $kT=0.2$~MeV
and $\tau_{\mathrm{NSE}} \sim 10^{-9}$ seconds for $kT=1$~MeV.
 In the core of a typical pre-supernova star
with $\rho=10^9$~g/cm$^3$ and $kT=0.32$~MeV we have $\tau_{\mathrm{NSE}} \simeq 2$~days. 
A typical duration
of the Si burning stages depends on stellar mass and varies from few hours to 3 weeks.
During the thermonuclear explosion of type Ia supernova in the flame region temperatures 
grow up to $kT=0.4\ldots0.6$~MeV, the timescale $\tau_{\mathrm{NSE}} \simeq 5$~milliseconds, 
and the explosion time is of the order of 1~second.

The weak transmutation rate between protons and neutrons is denoted by $\dot{Y_e}$,
\begin{equation}
\label{Ye_dot}
\dot{Y_e} \equiv \frac{d Y_e(t)}{dt} = \lambda_{\nu_e} - \lambda_{\bar{\nu}_e},
\end{equation}
where:
$$
\lambda_{\nu} = \sum_k \lambda_{\nu}^{(k)} \frac{X_k}{A_k}, \quad \lambda_{\nu}^{(k)}  = \int_0^\infty \psi_k( \mathcal{E}_{\nu} )\; d \mathcal{E}_{\nu}.
$$ 
If the hydrodynamic timescale is longer than $\tau_{\mathrm{NSE}}$ and $\dot{Y_e}$
change slowly\footnote{Slow in the sense of eq.~\eqref{Ye_dot}, not actual weak rates $\lambda_{\nu}, \lambda_{\bar{\nu}_e}$,
which may be very high.} then we can safely assume a quasistatic
evolution in the three-parameter space.
Usually\footnote{As noted by \cite{Iwo},
relativistically invariant triad $T-n_B-Y_e$ where $n_B$ is conserved
baryon number density may be used if General Relativity formulation is required.} 
these parameters are temperature
$kT$, the density $\rho$ and the electron fraction $Y_e$. For the given
triad $(kT, \rho, Y_e)$ we are able to determine abundances of
all nuclei. This approximation is widely used in ''iron'' cores of pre-supernova stars, supernovae, 
nuclear networks, thermonuclear flames and nucleosynthesis studies. Under NSE  conditions the neutrino
emission is not much different from thermal processes (especially
if $\dot{Y_e}=0$), and no prior
knowledge of abundances is required. This allows e.g. for post-processing of
models with a known history of temperature, density and electron fraction.
If $Y_e(t)$ is not known we still can use the NSE approximation assuming
some value, e.g. $Y_e=0.5$ for symmetric nuclear matter. 
The composition (and therefore
neutrino emission) is extremely sensitive to small changes in $Y_e$
in the most interesting range of $Y_e=0.35\ldots0.55$ and
relatively low temperatures of $kT<0.5$~MeV. 
One method to overcome this problem
is to use the so-called tracer particles built into simulation
to remember the thermodynamic history of matter. In the next step one then finds the history
of $Y_e$. Another application of the NSE neutrino emission, described in \cite{Iwo},
is the subgrid-scale model of nuclear flame energetics in thermonuclear supernovae.

This paper is organized as follows: in Sect.~\ref{weak} we 
discuss spectra of individual nuclei under conditions of high temperature and density.
We use solar $^7$Be neutrinos as an example.
In Sect.~\ref{spectra} we use NSE and get neutrino emissivities and energy spectra, using FFN 
\cite{FFN_I, FFN_II, FFN_III, FFN_IV} weak rates.
Final section comprises concluding remarks and a programme for future theoretical neutrino
astronomy (calculations of the neutrino spectra and so on). 

For details related to the implementation of NSE the reader is directed to the accompanying paper, 
that is submitted
to Atomic Data and Nuclear Data Tables \cite{ADNDT_AO}.

\section{Neutrino spectrum from $\beta$ processes \label{weak} in thermal bath}

Bahcall \cite{PhysRev.126.1143, PhysRev.128.1297} laid fundamental theoretical foundations
in the context of Solar neutrino spectrum. Later work
is upgrade for the results of Bahcall concerning the number of nuclei involved,
better nuclear data etc. With notable exception of the Sun \cite{1989neas.book.....B} 
and geo-neutrinos \cite{Enomoto_PhD}
a rigorous treatment of the neutrino spectra from individual nuclei
is usually ignored in astroparticle physics. Core-collapse simulations use parameterized
approach, cf. e.g. \cite{Bruenn1985, Langanke_neutrino_spectrum}. Unfortunately, in the case of multi-peaked
neutrino spectrum this approach simply does not work, 
cf. Fig.~1 and related comments in \cite{Langanke_neutrino_spectrum}. The antineutrino
spectrum is computed only for free neutrons, in applications known to the author.

The spectrum of neutrinos emitted from single nuclei in astrophysical
plasma depends strongly on the temperature and the chemical potential
of electrons (and positrons if $kT \sim m_e=0.511$~MeV or larger).
The temperature is large in typical evolutionary advanced astrophysical objects
(pre-supernova or supernova, for example). We will study neutrino spectrum in this regime.
On the contrary, for the solar interior, $kT=1.35 \times 10^{-3}, \mu_e=0$, and this makes
little change with respect to laboratory experiments.

Let us begin with typical example of the continuum electron capture process:
$$
^7\mathrm{Be} + e^- \to ^7\mathrm{Li} + \nu_e
$$

We make assumptions concerning the infinite nucleus mass and we
neglect various correction factors (screening, Coulomb factor). 
Then the $\epsilon^\pm$ capture rate is proportional to the constant matrix element
multiplied by the so-called phase space factor $\Phi$:
  \begin{equation}
    \label{capture} 
    \Phi_c
=  
        \frac{ 
      \mathcal{E}_{\nu}^2 ( \mathcal{E}_{\nu} - \Delta Q )
      \sqrt{ (\mathcal{E}_{\nu}-\Delta Q)^2 - {m_e}^2}
           }
           { 
            1+\exp{[ (\mathcal{E}_{\nu}-\Delta Q - \mu)/kT  ]}
           }    \Theta(\mathcal{E}_{\nu} - \Delta Q  - m_e),
  \end{equation}
where $\mathcal{E}_{\nu}$ denotes the neutrino energy for $e^-$
capture and $\mathcal{E}_{\nu}$ is the antineutrino energy for $e^+$ capture.
$\Delta Q$ is the energy difference between initial and final states
(both can be excited) and $m_e$ is the electron rest mass.
The chemical potential $\mu_e$ of the electron includes $m_e$, and therefore
for positrons $\mu_{e^+}=-\mu_{e^-} \equiv -\mu$; $kT$ is the temperature
of the electron gas.

It is worth to notice, that by expressing factor \eqref{capture} by 
the neutrino (antineutrino) energy rather than electron (positron) energy, we have 
just one formula, since both signs of $\Delta Q+m_e$ are covered, 
and $\mathcal{E}_{\nu}>0$.

The neutrino spectrum from $\beta^\pm$ decay is proportional to:
  \begin{equation}
    \label{decay}
    \Phi_d
=
      \frac{ 
      \mathcal{E}_{\nu}^2 (\Delta Q - \mathcal{E}_{\nu})
      \sqrt{ (\mathcal{E}_{\nu}-\Delta Q)^2 - {m_e}^2}
           }
           { 
            1+\exp{ (\mathcal{E}_{\nu}-\Delta Q + \mu)/kT  }
           }    \Theta(\Delta Q  -m_e - \mathcal{E}_{\nu})
  \end{equation}

\begin{figure}
\includegraphics[width=0.5\textwidth]{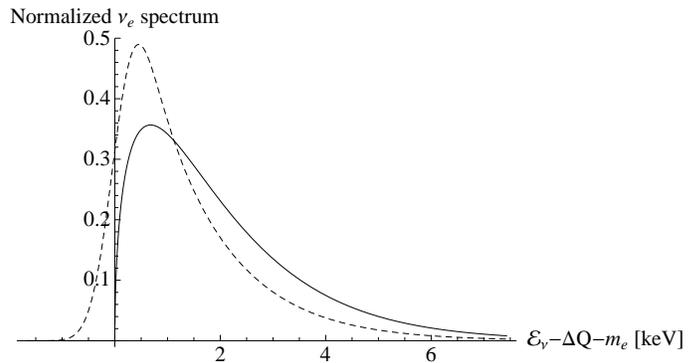}
\caption{ \label{Be7-norm}
The normalized neutrino spectrum for solar $^7$Be 
electron capture neutrino line, computed according to \eqref{capture} (solid line)
and the state-of-art result computed by Bahcall ( \cite{Bahcall-Be7-line}, Eq.~(46) )  
is shown by dashed line. 
}
\end{figure}

\begin{figure*}
\includegraphics[width=\textwidth]{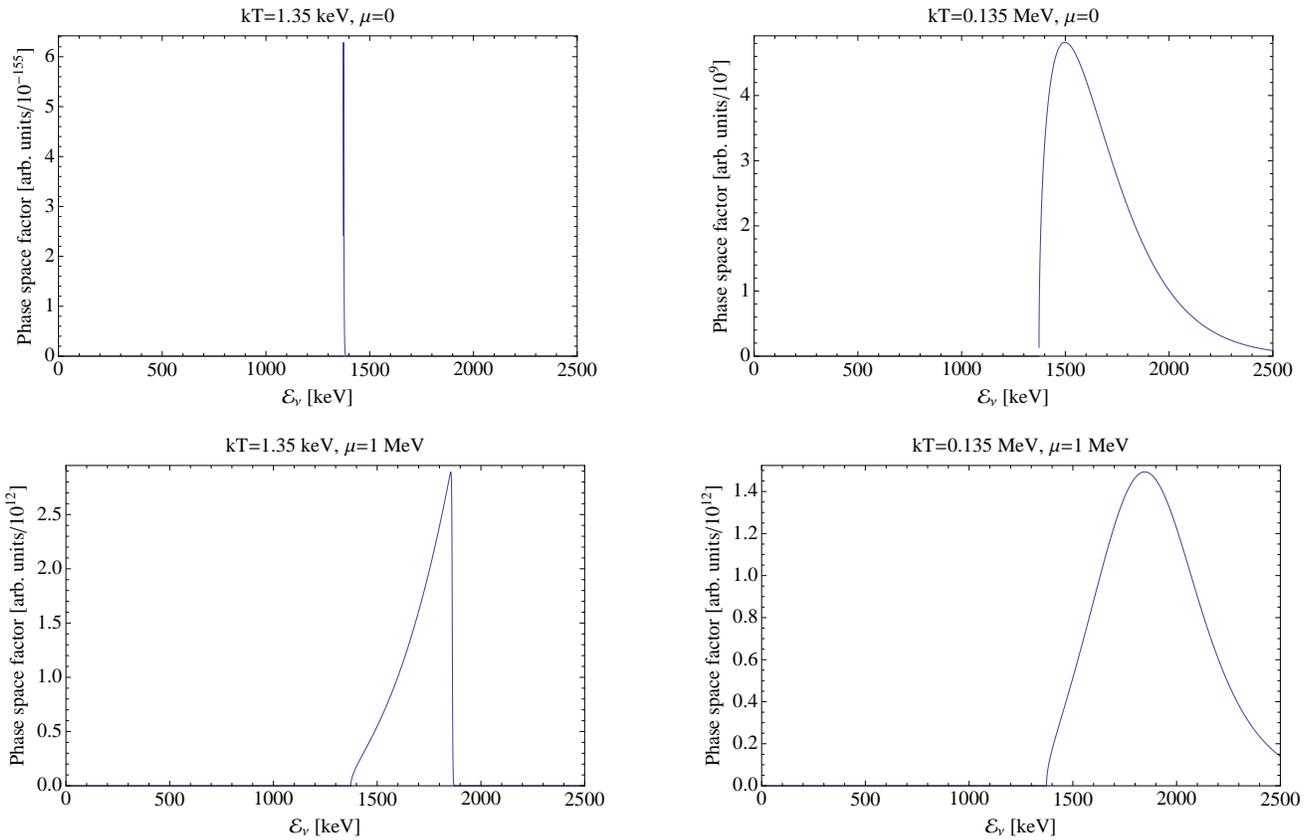}
\caption{ \label{Be7-4cases}
  The influence of the degeneracy (large $\mu$) and high
  temperatures (large $kT$) on the electron capture neutrino spectrum. Upper-left figure is for
  solar neutrinos (in laboratory conditions) and lower-left for cold degenerate electron gas. The upper-right
  figure refers to high temperature and the lower-right figure includes both degeneracy and high temperature.
}
\end{figure*}

Figure \ref{Be7-norm} compares neutrino spectrum given by formula \eqref{capture}
with the more elaborated result of \cite{Bahcall-Be7-line} for solar neutrinos.
Results are in good qualitative agreement. In both cases shown in Fig.~\ref{Be7-norm}
the neutrino spectrum is simply a line of negligible ( Fig.~\ref{Be7-4cases}, upper-left ) width.
The horizontal axis in Fig.~\ref{Be7-norm} is the difference between the Q-value (including $m_e$) and
the neutrino energy, in keV. This is because for solar conditions the Q-value for $^7$Be capture
is by many orders of magnitude larger than the temperature and the chemical potential
of the electron gas. If we put $^7$Be into a plasma where $kT$ or $\mu$
is comparable to the Q-value, both capture rate and neutrino spectrum change
dramatically, cf. Fig~\ref{Be7-4cases}. In general, spectrum shape
is a result of the competition between the Fermi-Dirac distribution
and the unit step function $\Theta$ in \eqref{capture}. While $e^-$ kinetic energy
always adds to the neutrino energy, for low 
temperatures it is negligible compared to $\Delta Q+m_e$.
If the temperature becomes non-negligible compared to Q-values, say $kT>0.1$~MeV, the thermal
broadening due to kinetic energy of electrons becomes important and
the capture rate is enhanced, cf. Fig.~\ref{Be7-4cases}, upper-right panel. 
For some of the laboratory stable nuclei the electron (positron) capture might 
be possible for high energy electrons (positrons), from thermal distribution tail.

The increase of density result in large $\mu_e$, and that leads to a more
visible effect. This is because most of the electrons, not just a small fraction
from the tail, have large energies. The neutrino spectrum (Fig.~\ref{Be7-4cases}, lower-left)
has a very characteristic shape in this case, with sharp edge on the high $\mathcal{E}_{\nu}$ end.
With the increasing $\mu_e$ progressively more nuclei become
unstable to the electron capture with the continuously growing capture rate. Lower-right panel in Fig.~\ref{Be7-norm}
shows effect of large $kT$ and $\mu$.

Anyway, possibly the most striking feature of Fig.~\ref{Be7-4cases} is not the shape of the spectrum 
but the dramating scale change on the vertical axis. Weak rates are extremely sensitive to both
$kT$ and $\mu$, mainly due to phase-space factors (\ref{capture}, \ref{decay}).

      \begin{figure}
      \includegraphics[width=0.5\textwidth]{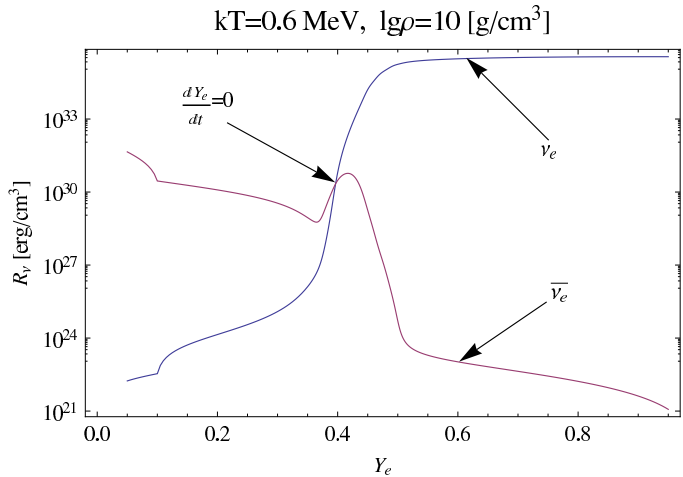}
      \caption{
      "(Color online)"
      \label{nue-nuebar} Neutrino and antineutrino emissivities as functions of $Y_e$. 
      Critical $Y_e$, defined as $\dot{Y_e}=0$, is seen at the crossing
      point of the neutrino and the antineutrino particle emission rates.}
      \end{figure}

\begin{figure}
      \includegraphics[width=0.5\textwidth]{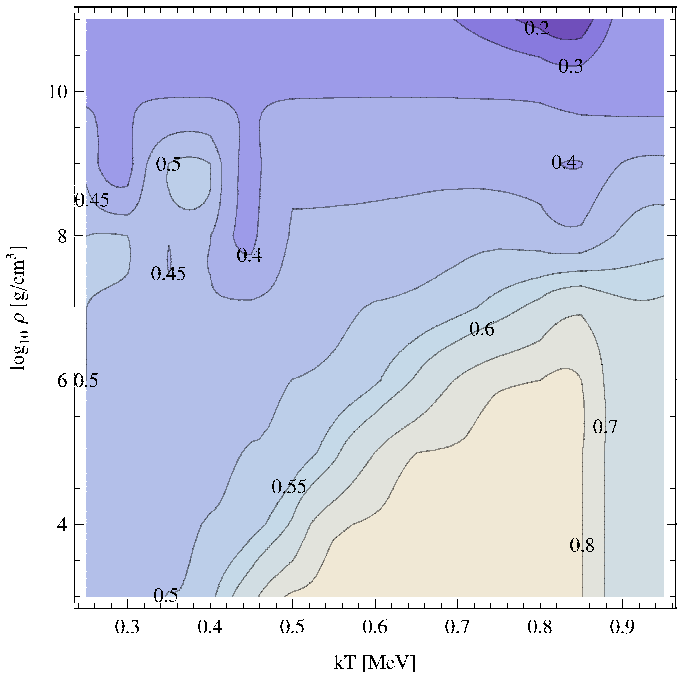}
      \caption{
      "(Color online)"
      \label{Ye-critical} Critical $Y_e$ (see Fig.~\ref{nue-nuebar} for explanation) for
       a range of considered temperatures and densities.}
      \end{figure}

In order to  get combined NSE spectrum we have to sum up all terms (\ref{capture},\ref{decay})
for all relevant pairs of excited states, multiply them by the partition function and matrix elements, 
and then substitute
into Eq.~\eqref{Xk} with $X_k$ obtained from NSE \cite{ADNDT_AO}.
A typical behavior of the NSE $\nu_e$ and $\bar{\nu}_e$ emissivities  \cite{FFN_I, FFN_II, FFN_III, Langanke_1, Langanke_2, Nabi_1} 
as a function of $Y_e$ is presented in Fig.~\ref{nue-nuebar}. 
As $Y_e$ decrease, electron neutrino flux (produced mainly in electron captures on protons
and by heavy nuclei) also tends to decrease. On the other hand, a decrease in  $Y_e$ cause
an increase in the flux of $\bar{\nu}_e$'s. Usually antineutrino emissivity peaks
due to beta decays of heavy nuclei and rise due to the positron capture
on neutrons and by neutron decay, cf. Fig.~\ref{nue-nuebar}. 
For almost all pairs ($kT$, $\rho$) we can find 
the value of $Y_e$ (Fig.~\ref{Ye-critical})  
where the flux of $\nu_e$ is equal to the flux of $\bar{\nu}_e$. These threshold values are particularly
interesting for the neutrino astronomy, because they might lead to the strong neutrino and antineutrino 
emission without further neutronization, in agreement with constraints from nucleosynthesis. 
The increase of the  $\bar{\nu}_e$ flux (with decreasing $Y_e$) 
stops neutronization a little bit earlier than derived from e.g. the expansion of matter and
the related decrease in rates alone.  The neutronization can also stop if $Y_e$ becomes too 
low and positron captures/$\beta^-$ decays start to dominate. 
Surprisingly, these critical $Y_e$ values (defined as $Y_e$ 
for which $\lambda_{\nu_e}=\lambda_{\bar{\nu}_e}$, Fig.~\ref{nue-nuebar} and Eq.~\eqref{Ye_dot}) vary
in a broad range (Fig.~\ref{Ye-critical}), reaching values close to $Y_e=0.875$ (primoidal BBN mixture 
of hydrogen and helium) for low densities and $kT=0.5\ldots0.8$. On the other hand, 
for highest densities ( $\rho>10^{11}$ g/cm$^3$) and temperatures kT$\simeq$0.8~MeV
an equilibrium sets at $Y_e=0.2\ldots0.3$. It is important to notice, that due to the low accuracy 
of the weak rates
derived from FFN tables and the variability of the NSE state with $Y_e$, 
Figure~\ref{Ye-critical} provides only a very approximate outlook
of critical values. 
The critical value\footnote{The state with $\dot{Y_e}=0$ is frequently refereed to as 
\em{kinetic beta equilibrium}. } 
is also very important for NSE timescales, as ''stalled'' $Y_e$ provide additional
time without breaking assumption on the quasistatic $Y_e$ evolution.  

      The competition between $\nu_e$ and $\bar{\nu}_e$ emission (Fig.~\ref{Ye-critical}) 
      is usually described in terms of the balance between
      electron captures (mainly on protons) and $\beta^-$ decays of the heavy nuclei \cite{Aufderheide}.
      However, for $Y_e$ outside range of $0.35..0.45$ the most important process leading to the 
      $\bar{\nu}_e$ emission is the positron capture by neutrons.

\section{Spectra under astrophysical conditions of interest \label{spectra}}

\begin{figure*}
\includegraphics[width=\linewidth]{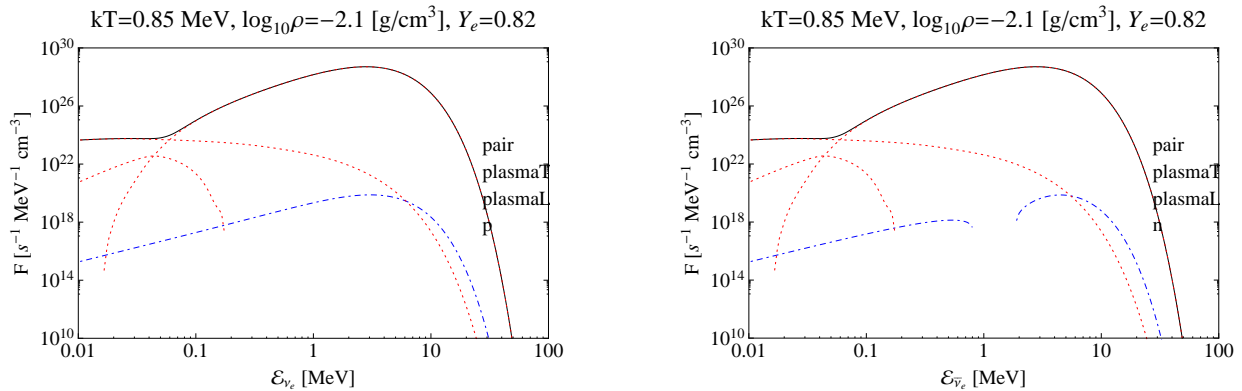}
\caption{
"(Color online)"
Neutrino (left) and antineutrino (right) spectrum emitted per unit volume
under conditions typical for BigBang nucleosynthesis era. 
\label{BBN_neutrinos}
}
\end{figure*}

\begin{figure*}
\includegraphics[width=\textwidth]{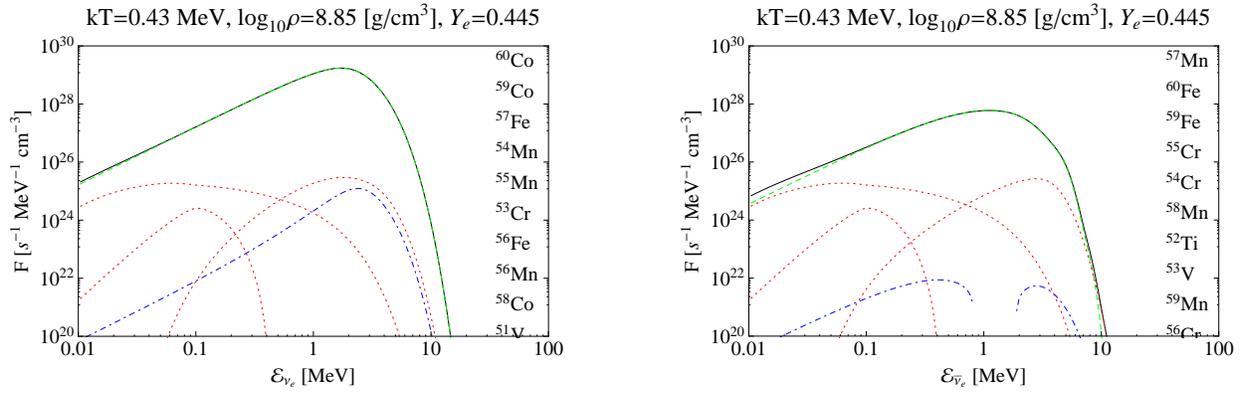}
\caption{
"(Color online)"
Neutrino (left) and antineutrino (right) spectrum emitted per unit volume
under conditions typical for presupernova star. 
\label{presn_neutrinos}
}
\end{figure*}

\begin{figure*}
\includegraphics[width=\linewidth]{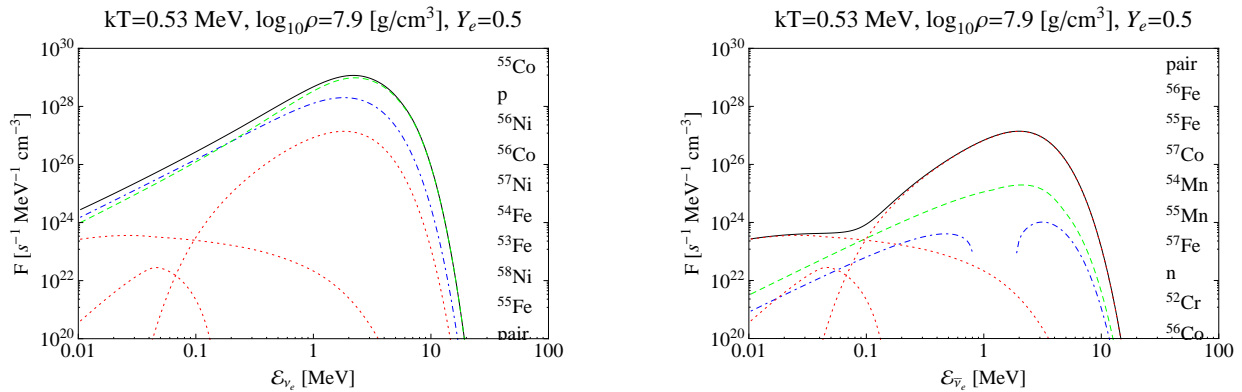}
\caption{
"(Color online)"
Neutrino (left) and antineutrino (right) spectrum emitted per unit volume
under conditions typical for detonation stage of thermonuclear supernova
in delayed-detonation class of models. 
\label{y12_neutrinos}
}
\end{figure*}

\begin{figure*}
\includegraphics[width=\linewidth]{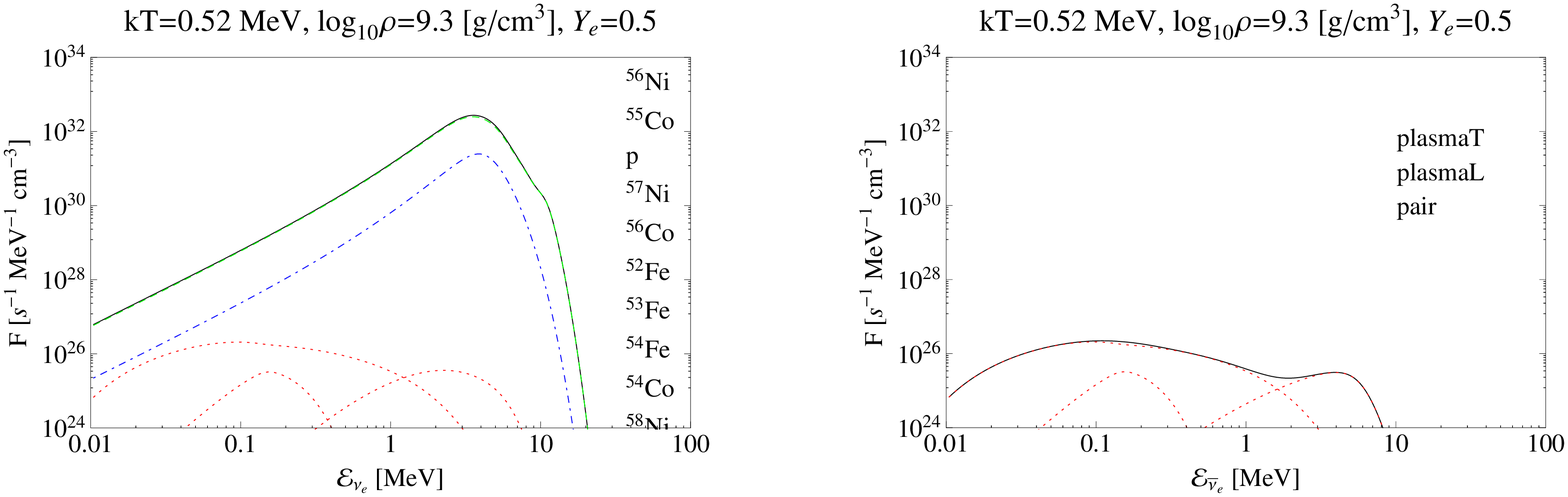}
\caption{
"(Color online)"
Neutrino (left) and antineutrino (right) spectrum emitted per unit volume
under conditions typical for early deflagration stage of thermonuclear supernova.
\label{n7_neutrinos}
}
\end{figure*}

\begin{figure*}
\includegraphics[width=\textwidth]{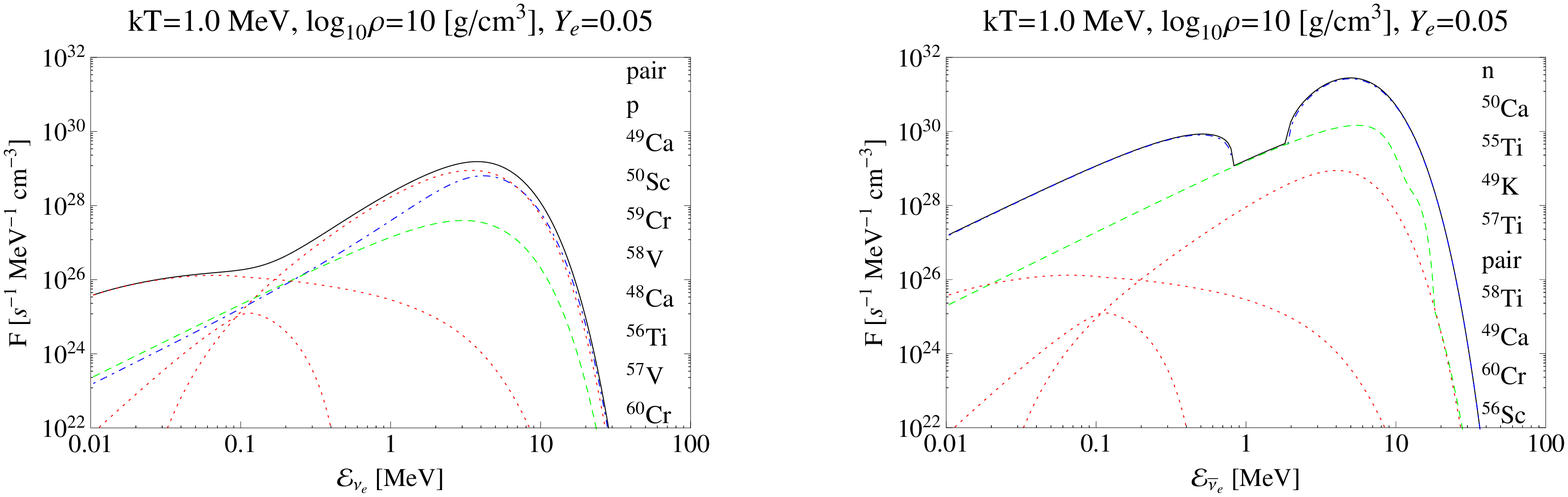}
\caption{
"(Color online)"
Neutrino (left) and antineutrino (right) spectrum emitted per unit volume
under conditions typical for neutron star merger. 
\label{merger_neutrinos}
}
\end{figure*}

\begin{figure*}
\includegraphics[width=\textwidth]{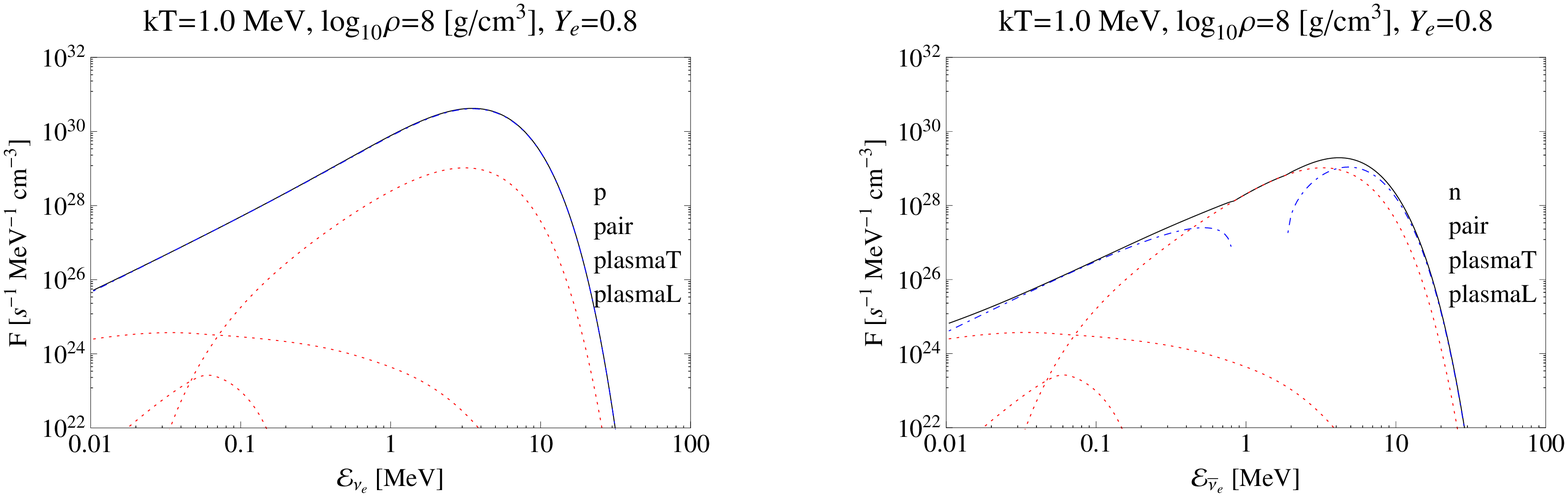}
\caption{
"(Color online)"
Neutrino (left) and antineutrino (right) spectrum emitted per unit volume
under conditions of large $Y_e$ and high temperature. 
\label{lsd_neutrinos}
}
\end{figure*}

\begin{figure*}
\includegraphics[width=\textwidth]{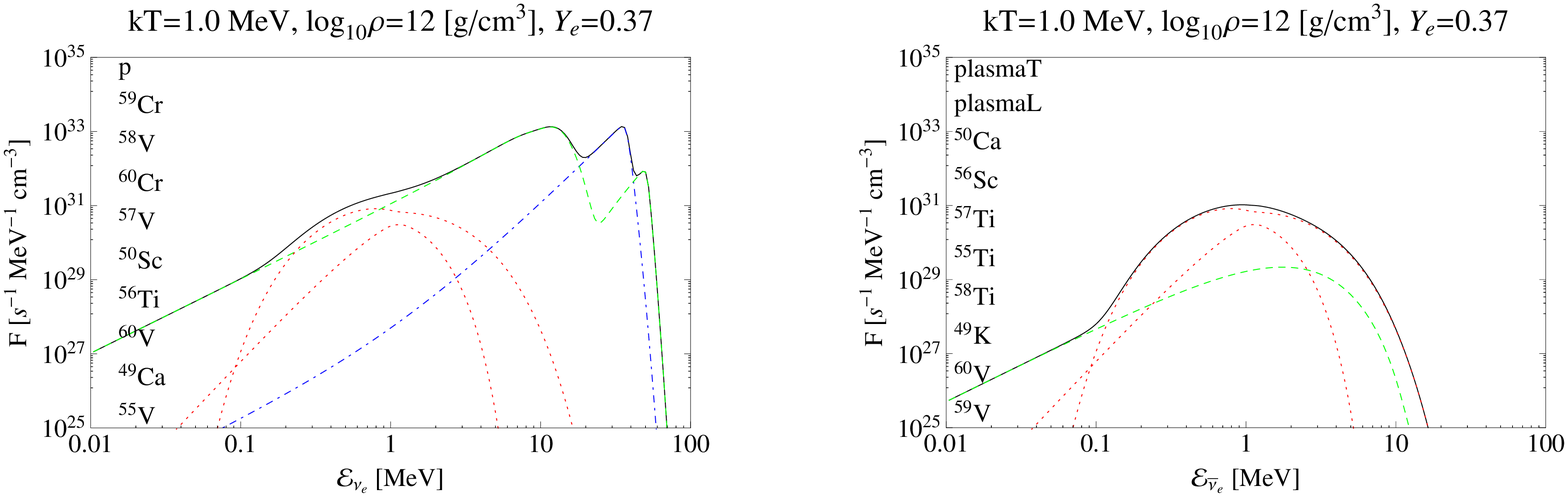}
\caption{
"(Color online)"
Neutrino (left) and antineutrino (right) spectrum emitted per unit volume
under conditions typical for the infall stage of the core-collapse. 
\label{cc_neutrinos}
}
\end{figure*}

We are able to compute approximate neutrino/antineutrino spectra for a wide range of astrophysical
phenomena if the  NSE timescale is short compared to dynamic and weak timescales.
Main limitation of our method is the neutrino trapping. 
The core-collapse supernovae and related phenomena,
e.g. long gamma-ray bursts, are examples of objects where neutrino trapping
is essential. We can use our method for initial infall stage of the collapse only.
But as long as we are inthe free streaming
regime this is the method of choice. We can produce much more detailed and accurate
neutrino spectra (than hydrodynamic simulations itself) {\em via} postprocessing.  
The latter is trivial to parallelize, and allow to achieve greater accuracy.
Our method can be applied to
cosmological-like \cite{Arnett} neutrinos (Fig.~\ref{BBN_neutrinos}), the center of 
the pre-supernova \cite{MassiveStars_rev1} star 
(Fig.~\ref{presn_neutrinos}) and typical conditions during type Ia thermonuclear explosions 
(Fig.~\ref{y12_neutrinos}, \ref{n7_neutrinos}). Other examples,
less interesting since there are many known results 
\cite{Shock_track, 1989A&A...224...49J, 2006ApJ...644.1028P, 2005ApJ...623..325P, 1994A&A...290..496J,
2007PhR...442...38J, 2003PhRvL..90x1101B, 1998ApJ...495..911M, 2006A&A...447.1049B, 2006A&A...457..963S,
2006A&A...457..281B, 2005ApJ...620..840L, 2000ApJ...539L..33R, 2007PhRvL..98z1101O, 2002A&A...396..361R,
2000PhRvC..62c5802T, 2006ApJ...640..878B, 1986ApJ...307..178B, 1995ApJ...450..830B, 2007ApJ...664..416B, 
2003ApJ...592..434T, 2005ApJ...626..317W, 1988ApJ...334..891B, 2006NuPhA.777..356B, 2008ApJ...688.1159M, 
1993ApJ...405..669M, 2001PhRvL..86.1935M, 2004ApJS..150..263L, 2003PhRvL..91t1102H,
2003PhRvL..90x1102L}
are provided for the core-collapse SN (Fig.~\ref{cc_neutrinos}) and the
merger of NS (Fig.~\ref{merger_neutrinos}). The spectrum might be calculated for exotic conditions
which are not related to any recently considered model as well, cf. Fig.~\ref{lsd_neutrinos}.
Examples are listed in Table~\ref{tbl}.

\begin{table}
\caption{Examples of neutrino and antineutrino spectra
\label{tbl}
}
\begin{tabular}{c|ccc|cc}
Object       & kT [MeV] (T$_9$) & $\rho$ [g/cm$^3$]   & $Y_e$ & Figure & Refs.\\
\hline
\hline
BBN          & 0.85 (9.9)       & 0.008               & 0.82  & Fig.~\ref{BBN_neutrinos} &\cite{Arnett} \\
Pre-SN       & 0.43 (5.3)       & 7.0 $\times 10^8$   & 0.445 & Fig.~\ref{presn_neutrinos} &\cite{MassiveStars_rev1}  \\
SN Ia DET    & 0.53 (6.1)       & 7.8 $\times 10^7$   & 0.5   & Fig.~\ref{y12_neutrinos} &\cite{Plewa_1, Plewa_2}  \\
SN Ia DEF    & 0.52 (6.0)       & 2.0 $\times 10^9$   & 0.5   & Fig.~\ref{n7_neutrinos} &\cite{Plewa_1, Plewa_2}  \\
NS-NS merger & 1.0 (11.6)       & 1.0 $\times 10^{10}$& 0.05  & Fig.~\ref{merger_neutrinos} &  \cite{rr}        \\
-            & 0.9 (11.6)       & 2.0 $\times 10^9$   & 0.8   & Fig.~\ref{lsd_neutrinos} & - \\ 
CC SN\footnote{Note: this example pushes our method to the limits of applicability.
More realistic spectrum is different, because neutrinos are trapped and they begin to diffuse rather than
escape freely.
}       
             & 1.0 (11.6)       & 1.0 $\times 10^{12}$& 0.73   & Fig.~\ref{cc_neutrinos} &\cite{Arnett}
\end{tabular}
\end{table}

Illustrative example is provided using cosmological weak freezout values as input. Following
\cite{Arnett} we put $T_9 \simeq9.9$, $\rho=0.008$~g/cm$^3$ and $Y_e=0.82$.
Neutrino and antineutrino spectrum in Fig.~\ref{BBN_neutrinos} is produced mainly
from pair annihilation process. Therefore, spectrum is almost purely thermal. This is not
a surprise, because thermal spectrum is what is expected for Big Bang neutrinos. Thus, our method is working
qualitatively well even in this extremal example. 

Pre-supernova stars are neutrino sources of particular interest \cite{OMK1}. The neutrino spectrum has been
obtained using values at the center of a star  during maximum compression stage. This stage is achieved 
just prior to the shell Si ignition
above iron core, few hours before the start of the collapse. Spectrum for $kT=0.43$~MeV, 
$\rho=7 \times 10^8$~g/cm$^3$ and $Y_e=0.445$ is presented 
in Fig.~\ref{presn_neutrinos}. The important reference for these numbers is \cite{MassiveStars_rev1}.
A striking feature in Fig.~\ref{presn_neutrinos} is the significant contribution from heavy nuclei for both
$\nu_e$ and $\bar{\nu}_e$ spectra. This is particularly important for the detection of these neutrinos,
as previous studies \cite{OMK1} were based solely on the thermal emission. For $\bar{\nu}_e$ 
the pair annihilation process dominates the high energy tail ($\mathcal{E}_{\bar{\nu}_e}>10$ MeV),
and the number of detectable inverse-$\beta$ events in a standard large water Cherenkov detectors 
\cite{OMK1, OMK2, Odrzywolek_SN1987A-20th}  
does not change. For a different detector design, e.g. a liquid scintillator \cite{LAGUNA}
the threshold might be as low as 0.2 MeV \cite{2008arXiv0806.2400B} and the number of events will be much 
larger than anticipated from thermal processes only. The situation is even more pronounced for electron neutrinos. 
A large number
of nuclei participate in massive electron captures leading to the flux that is two orders of magnitude larger 
than from pair process even for  $\mathcal{E}_{\nu_e}>5$ MeV. Therefore, the detection of $\nu_e$'s, 
previously rejected from analysis due to experimental difficulties, should be reconsidered.
The contribution from free nucleons can be neglected in pre-supernova case.

Another very important example of application for our method is the type Ia thermonuclear
supernova.
Two important regimes for thermonuclear burning in type Ia supernovae are deflagration
and detonation. For deflagration we use $kT=0.53$~MeV and $\rho=2 \times 10^9$ g/cm$^3$. For detonation
a similar temperature of $kT=0.52$~MeV has been used, but the density has been reduced due to pre-expansion to 
$\rho=7.8 \times 10^7$ g/cm$^3$. $Y_e=0.5$ was used, but it is important to point out that small
neutronization is inevitable, because the $\nu_e$ flux dominates over the flux of $\bar{\nu}_e$, cf.~Fig.~\ref{n7_neutrinos}. Neutrino emission is very sensitive to these small changes. 
Model y12 and n7d1r10t15c of \cite{Plewa_1, Plewa_2} are the sources of the values used above.
For a type Ia supernovae free nucleons are among the top neutrino sources, see 
Figs.~\ref{y12_neutrinos}, \ref{n7_neutrinos}. Two presented cases are related to 
detonation and deflagration. A lower density used for detonation
stage (Fig.~\ref{y12_neutrinos}) is the result of the white dwarf pre-expansion due to the previous
deflagration stage \cite{Plewa_1}. The high density in Fig.~\ref{n7_neutrinos} is connected
to the initial stage of subsonic nuclear burning in the pure deflagration model of \cite{Plewa_1}.
Only three nuclei contribute
significantly to the $\nu_e$ spectrum in both cases: $^{55}$Co, $^{56}$Ni and protons.
Relative contributions and total flux are different, however. The neutrino flux per unit volume
is four orders of magnitude larger for deflagration compared to detonation. Nevertheless,
the deflagration involve tiny volume of the white dwarf only, while the detonation wave usually 
traverses entire star. Integrated flux might be similar, but this is model-dependent. 
Antineutrino spectrum is dominated by thermal processes: pair annihilation
during detonation and plasmon decay during deflagration. The total flux is much smaller 
than for neutrinos, and this imbalance causes $Y_e$ to decrease. Therefore results from
Figs.~\ref{y12_neutrinos} and \ref{n7_neutrinos}, with assumed $Y_e=0.5$, should be taken with
care. For example, the NSE abundance of $^{55}$Co drops rapidly in the range $Y_e=0.5\ldots0.47$.
A more detailed investigation of type Ia neutrinos shows also an important contribution
from free neutrons to the anti-neutrino spectrum above 2 MeV.

Now, we study neutrino spectrum for the accretion disk formed in NS-NS merger.
Needed data are taken from Fig.~1 of \cite{rr}:
temperature of $kT=1.0$ MeV, $\rho=10^{10}$ g/cm$^3$ and $Y_e=0.05$.
Similar results are
expected for the neutron star - black hole mergers and other phenomena forming low $Y_e$,
dense, high temperature accretion disks. The neutrino spectrum is a result of pair process
and electron captures on protons. The antineutrino  spectrum is heavily dominated by the 
neutron decay and positron captures on neutrons. The gap $0.8 < \mathcal{E}_{\bar{\nu}_e} < 1.8$ MeV
is filled by processes involving heavy nuclei. Moreover, antineutrino flux is much larger
compared to the neutrino flux. The spectrum peaks at $\mathcal{E}_{ \bar{\nu}_e } \simeq 5$~MeV,
providing interesting candidate for the neutrino detection using the inverse $\beta$ decay.

Another example (Fig.~\ref{lsd_neutrinos}), not related to a particular astrophysical phenomena, shows
the importance of thermal processes and those involving free nucleons.
The antineutrino spectrum, especially the high energy end due to the positron capture
is particularly important. Spectral features of this process should be interesting 
for future neutrino astronomy, based on gigantic
$\bar{\nu}_e$ water-based detectors \cite{0901.1950, abe-2007, debellefon-2006, LAGUNA}.

The core-collapse process is poorly described using our method, but we have provided an example
for the sake of the completeness. The calculated neutrino spectrum in Fig.~\ref{cc_neutrinos} has a complex multi-peak structure. 
This is in contrast
to the results of the more sophisticated neutrino radiation transport results, which are
always single-peaked. This can be explained
by: (1) smoothing nature of the diffusive transport and (2) too small energy resolution (to few energy bins)
of the transport codes used in simulations. The high energy neutrinos seen in Fig.~\ref{cc_neutrinos} 
are in reality downscattered
to much smaller energies. The same applies to antineutrinos. Additionally, $\nu$-$\bar{\nu}$ pairs are created
in the process of collisions between neutrinos and electrons, and between pairs of neutrinos 
 \cite{2003ApJ...587..320B}. 
This leads to the energy exchange between flavors, and realistic $\nu_x$ spectra\footnote{Muon and tau spectra
are almost identical to the thermal electron flavor spectra, except for smaller integrated flux.} 
are not as distinct as those
from Figs.~\ref{cc_neutrinos}. Factors that block outgoing neutrinos
and could shape the neutrino spectrum under such extreme
conditions were omitted.
Clearly,
our method is not working for the core-collapse supernovae, as anticipated.

\section{Conclusions}

One of our important conclusions is related to typical way of publishing data on weak nuclear processes 
in astrophysics. This approach dates back into year 1980, and was introduced
in the famous paper \cite{FFN_I}. Tables published by the FFN become standard
in modern astrophysics. Upgrades  \cite{FFN_I, Langanke_ADNDT, Nabi_1} did not change
structure of FFN tables. Unfortunately, FFN grid using mere 13x11 points is not 
enough to obtain precise results, as noted already by the FFN authors \cite{FFN_IV}. 
While we understand reasons to preserve this standard for 30 years, "reverse engineering"
of FFN-like tables to get spectrum, as well as complicated interpolating 
procedure is impractical now.
If one wants to calculate the spectrum precisely, without analytical approximate formula for individual nuclei,
pre-calculated tables are useless. Much more convenient is the following set of data: 
\begin{enumerate}
    \item energy and spins for ground and excited states
    \item weak transition matrix elements between all relevant pairs
          of the excited states for the parent and daughter nuclei
\end{enumerate} 
Alternatively, tabulated spectrum for all $T-\rho Y_e$ pairs would be a good choice, with
amount of stored data up to several megabytes.
While such approach will increase amount of published numerical data by a 
factor of $\sim10$, it would remove any ambiguity in the representation of the spectra.

The inspection of virtually any of the figures presented here (Figs.~\ref{BBN_neutrinos}-\ref{cc_neutrinos}) 
clearly show the importance of both nuclear and thermal processes.
The thermal emission and captures on free nucleons and nuclei should be included in consistent calculations. However, 
depending on the subject, all combinations of these can be found in astrophysical applications. For example, 
type Ia supernova simulations include NSE emission but older simulations neglect neutrino emission at all or
include electron captures only \cite{KunugiseIwamoto}. Other important regimes, core-collapse and pre-supernovae frequently neglect
positron captures, particularly on neutrons. Estimates of the neutrino signal in detectors 
from pre-supernovae rely purely on thermal emission \cite{OMK1, OMK2, Odrzywolek_SN1987A-20th}.

The ultimate goal which is beyond scope of the article is to know exactly (not approximately!)
the neutrino spectrum  from weak nuclear processes under NSE. In the past weak rates were usually
integrated and only the total neutrino flux (particles and energy) 
has been tabulated and presented to the public. We argue again,
that this is not the best approach if one wants to calculate 
the neutrino spectrum. Without full input used to calculate weak rates
we are unable to restore information lost in the integration. 
Typical (FFN-like) weak interaction tables are not sufficient.
Tables of the excited states, spins and weak matrix elements
for all considered nuclei will allow researchers to calculate
both neutrino/antineutrino spectra and customized weak interaction
rate tables.

Weak rates prepared in the FFN fashion (i.e. all published rates \cite{FFN_I, Langanke_ADNDT, Nabi_1}), 
even those with tabulated effective $\lg{ \langle ft \rangle}$, do not facilitate estimates of the neutrino spectrum.
This is not surprising, because these rates were prepared for a different purpose: the neutrino energy loss and
neutronization. Maximal information on the spectrum extracted from FFN-like tables can be extracted as described 
in the paper accompanying paper \cite{ADNDT_AO}.
We re-tabulate effective $\lg{ \langle ft \rangle}$ values and effective $Q_{eff}$-values for every grid point
to get from \eqref{capture} or \eqref{decay} the original total rate and average neutrino energy. 
If the total rate is not dominated by the captures we switch from \eqref{capture} to \eqref{decay}. This approach produces significant 
side effects if capture and decay rates are comparable. The neutron provides good example. Due to the non-negligible contribution 
of $\bar{\nu}_e$'s from the neutron decay, the average energy differ from that deduced from pure positron capture.
Therefore the effective spectrum has a variable effective Q-value. The realistic positron (and electron as well) capture spectrum 
always starts with energy equal to the lowest Q-value. To sum up, the obvious next step in the research is to give up pre-calculated
tables of weak rates and to re-calculate the neutrino spectrum from scratch, using nuclear data and weak matrix elements
as an input.

Despite these difficulties, we obtained new results. 
\begin{enumerate}
\item we get interpolating procedures for NSE abundances with number of convenient features: the ability to pick out of NSE 
        selected nuclei, the computational time scaling linearly with the number of nuclides and independent
        of the position in $T-\rho-Y_e$ space for full $Y_e=0.05\ldots0.95$ range \cite{ADNDT_AO} 
\item \textbf{the energy spectrum}, fluxes, mean energies etc.  of the emitted neutrinos and antineutrinos \textit{separately} for
      $\nu_e$ and $\bar{\nu_e}$
\end{enumerate}

Our analysis was meant to be general,  but we can identify some possible astrophysical targets for presented methods. 
The NSE neutrino spectrum would be a good approximation for massive stars after Si burning and thermonuclear supernovae. 
A related research is underway. Procedures developed here will be 
useful for the analysis of neutrino signals from X-ray flashes, neutron stars, merger events, accretion disks 
and some types of cosmic explosions, e.g. pair-instability supernovae.

The electron antineutrino emission due to the positron capture
on neutrons provides strong and relatively high-energy flux for surprisingly large volume 
in $kT - \rho - Y_e$ space.
Needed thermodynamic conditions: $kT>0.6$ and $\rho>10^7$ g/cm$^3$ can be met in many astrophysical objects. 
Megaton-scale neutrino detectors \cite{2008arXiv0810.1959K} will search for antineutrinos with 
energy $\mathcal{E}_{\bar{\nu_e}}>1.8$~MeV.
The detection of strong $\nu_e$-flux above 5~MeV produced mainly by captures on protons and heavy nuclei is 
standard in water Cherenkov \cite{0901.1950, abe-2007, debellefon-2006} or 
liquid scintillator \cite{LAGUNA, Beacom_PES, 0901.1950} detectors.
Therefore further investigation of NSE neutrinos, particularly in the unexplored region of large $0.87>Y_e \gg 0.55$ should
give researchers some additional hints for the existence (or non-existence)  of detectable astrophysical antineutrino sources. 
 
\begin{acknowledgments}
I would like to thank I.~Seitenzahl for discussion
 of the NSE calculations and T.~Plewa for motivation
and support of this work. My colleagues, E.~Malec and S. Dye, contributed
significantly to this work carefully reading
the manuscript. I also thank to anonymous referee
for important suggestions making presentation of the results
much more useful for astrophysical community. 
\end{acknowledgments}

\bibliographystyle{apsrev}

\bibliography{NSE_neutrinos_part1}

\end{document}